\documentclass[twocolumn]{aastex62}

\usepackage[T1]{fontenc}
\usepackage{amsmath}

\newcommand{\package}[1]{\textsl{#1}}
\newcommand{\gaia}{\textsl{Gaia}}

\newcommand{\tractor}{\textsl{The Tractor}}
\newcommand{\msun}{\textrm{M}_\odot}
\newcommand{\kpc}{\textrm{kpc}}
\newcommand{\kms}{\ensuremath{\textrm{km}~\textrm{s}^{-1}}}
\newcommand{\kmskpc}{\ensuremath{\textrm{km}~\textrm{s}^{-1}~\textrm{kpc}^{-1}}}
\newcommand{\bs}[1]{\boldsymbol{#1}}
\newcommand{\masyr}{\ensuremath{\textrm{mas}~\textrm{yr}^{-1}}}

\newcommand{\given}{\,|\,}

\renewcommand{\tablename}{Table}
\usepackage{upgreek}

\shorttitle{pal~5's biggest fan}
\shortauthors{bonaca et al.}

\begin{document}\sloppy\sloppypar\raggedbottom\frenchspacing 

\title{Variations in the width, density, and direction of the Palomar~5 tidal tails}

\author[0000-0002-7846-9787]{Ana~Bonaca}
\affiliation{Center for Astrophysics | Harvard \& Smithsonian, 60 Garden Street, Cambridge, MA 02138, USA}
\email{ana.bonaca@cfa.harvard.edu}
\correspondingauthor{Ana Bonaca}

\author[0000-0003-0256-5446]{Sarah~Pearson}
\affiliation{Center for Computational Astrophysics, Flatiron Institute, 162 Fifth Avenue, NY 10010, USA}

\author[0000-0003-0872-7098]{Adrian~M.~Price-Whelan}
\affiliation{Center for Computational Astrophysics, Flatiron Institute, 162 Fifth Avenue, NY 10010, USA}
\affiliation{Department of Astrophysical Sciences, Princeton University, 4 Ivy Lane, Princeton, NJ 08544, USA}

\author{Arjun~Dey}
\affiliation{National Optical Astronomy Observatory, 950 North Cherry Avenue, Tucson, AZ 85719, USA}

\author{Marla~Geha}
\affiliation{Department of Astronomy, Yale University, New Haven, CT 06520, USA}

\author{Nitya~Kallivayalil}
\affiliation{Department of Astronomy, University of Virginia, 530 McCormick Road, Charlottesville, VA 22904, USA}

\author{John~Moustakas}
\affiliation{Department of Physics \& Astronomy, Siena College, NY 12211, USA}

\author{Ricardo~Mu\~ noz}
\affiliation{Departamento de Astronomia, Universidad de Chile, Camino del Observatorio 1515, Las Condes, Santiago, Chile}

\author{Adam~D.~Myers}
\affiliation{Department of Physics and Astronomy, University of Wyoming, Laramie, WY 82071, USA}

\author{David~J.~Schlegel}
\affiliation{Lawrence Berkeley National Laboratory, 1 Cyclotron Road, Berkeley, CA 94720, USA}

\author{Francisco~Valdes}
\affiliation{National Optical Astronomy Observatory, 950 North Cherry Avenue, Tucson, AZ 85719, USA}

\begin{abstract}\noindent 
Stars that escape globular clusters form tidal tails that are predominantly shaped by the global distribution of mass in the Galaxy, but also preserve a historical record of small-scale perturbations.
Using deep $grz$ photometry from DECaLS, we present highly probable members of the tidal tails associated with the disrupting globular cluster Palomar~5.
These data yield the cleanest view of a stellar stream beyond $\sim20\,\rm kpc$ and reveal: (1) a wide, low surface-brightness extension of the leading tail; (2) significant density variations along the stream; and (3) sharp changes in the direction of both the leading and the trailing tail.
In the fiducial Milky Way model, a rotating bar perturbs the Palomar~5 tails and can produce streams with similar width and density profiles to those observed.
However, the deviations of the stream track in this simple model do not match those observed in the Palomar~5 trailing tail, indicating the need for an additional source of perturbation.
These discoveries open up the possibility of measuring the population of perturbers in the Milky Way, including dark-matter subhalos, with an ensemble of stellar streams and deep photometry alone.
\end{abstract}


\section{Introduction}
\label{sec:intro}

Direct N-body simulations of globular clusters orbiting in a static galactic potential predict that the clusters continually lose stars through evaporation and tidal stripping \citep[e.g.,][]{Baumgardt:2003}.
Stars escape the cluster with a small relative velocity and thus form thin, kinematically cold streams \citep[e.g.,][]{Combes:1999}.
As a result, globular cluster streams are excellent tracers of the underlying tidal field, and under the assumption of a static potential, they constrain the enclosed mass within their current location \citep{Bonaca:2018}.
Nearby stellar streams have already been used to measure the mass and shape of the Milky Way halo \citep[e.g.,][]{Koposov:2010, Kupper:2015, Bovy:2016}.

However, streams are long-lived and witness the host galaxy evolve, including its gradual increase in mass, its rotating bar, and orbiting dark matter subhalos.
When simulated in more realistic environments that feature some of these events, the resulting streams are no longer thin, coherent structures \citep[e.g.,][]{Bonaca:2014, Ngan:2015, Price-Whelan:2016b}. Recently, \citet{Price-Whelan:2018} detected gaps and off-the-stream features in the GD-1 stellar stream that could be signatures of perturbation \citep{Bonaca:2018b}.
This discovery establishes stellar streams as a cosmological probe of dark matter on small scales.
However, streams can also be affected by baryonic perturbers, such as giant molecular clouds \citep{Amorisco:2016}, the Galactic bar \citep{Pearson:2017}, and spiral arms \citep{Banik:2019}.
Streams can even naturally develop features in their density profile during cluster disruption \citep[e.g.,][]{Kupper:2008, Just:2009}.
Additionally, stream debris can spread out rapidly in phase space if their progenitors are evolving on non-regular orbits \citep[e.g.,][]{Pearson:2015, Fardal:2015, Price-Whelan:2016}.
To infer the abundance of dark-matter subhalos from stream perturbations, we need to confirm the origin of stream perturbations first.

In this paper, we revisit the tidal tails of the Palomar~5 (Pal~5) globular cluster \citep{Odenkirchen:2001, Rockosi:2002}.
The Pal~5 stream features density variations not reproduced in a static model of the Milky Way, but neither their significance nor their origin have been established \citep{Carlberg:2012, Bernard:2016, Ibata:2016, Erkal:2017}.
Arguably, Pal~5 provides the best opportunity for disentangling different mechanisms that shape the stream as it has a surviving progenitor.
Such an endeavor, however, requires a robust map of the entire tidal debris.
Pal~5 is located too far from the Sun to enable efficient membership selection based on \gaia\ proper motions, while accurate mapping using the existing photometry is limited because the catalogs are either wide, but shallow \citep{Bernard:2016}, or deep, but narrow \citep{Ibata:2016}.
To confidently identify Pal~5 members over a wide area, we use deep, wide-field photometry from the DECam Legacy Survey (\S\ref{sec:data}).
In \S\ref{sec:densitymodel} we use the updated map of Pal~5 to quantify how the stream track, width and density vary along the stream.
We then explore how these properties change across Pal~5 models simulated in a range of Galactic potentials (\S\ref{sec:sim}) and conclude with a discussion of perturbers that jointly could have caused the observed Pal~5 features (\S\ref{sec:discussion}).

\section{Data}
\label{sec:data}

\begin{figure*}
\begin{center}
\includegraphics[width=0.83\textwidth]{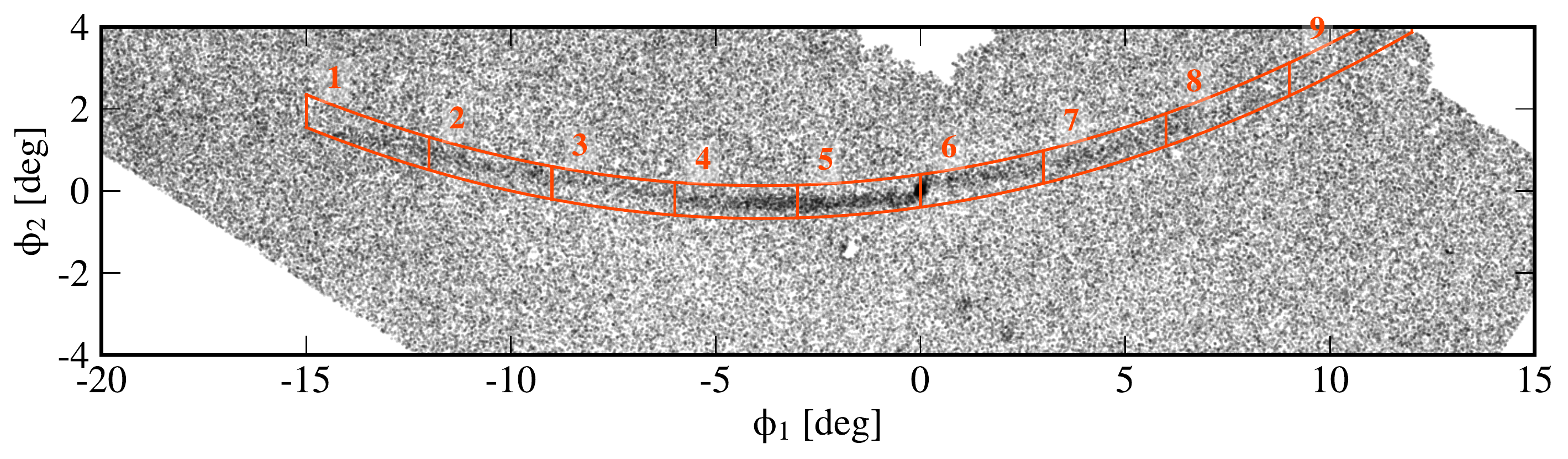}
\includegraphics[width=0.83\textwidth]{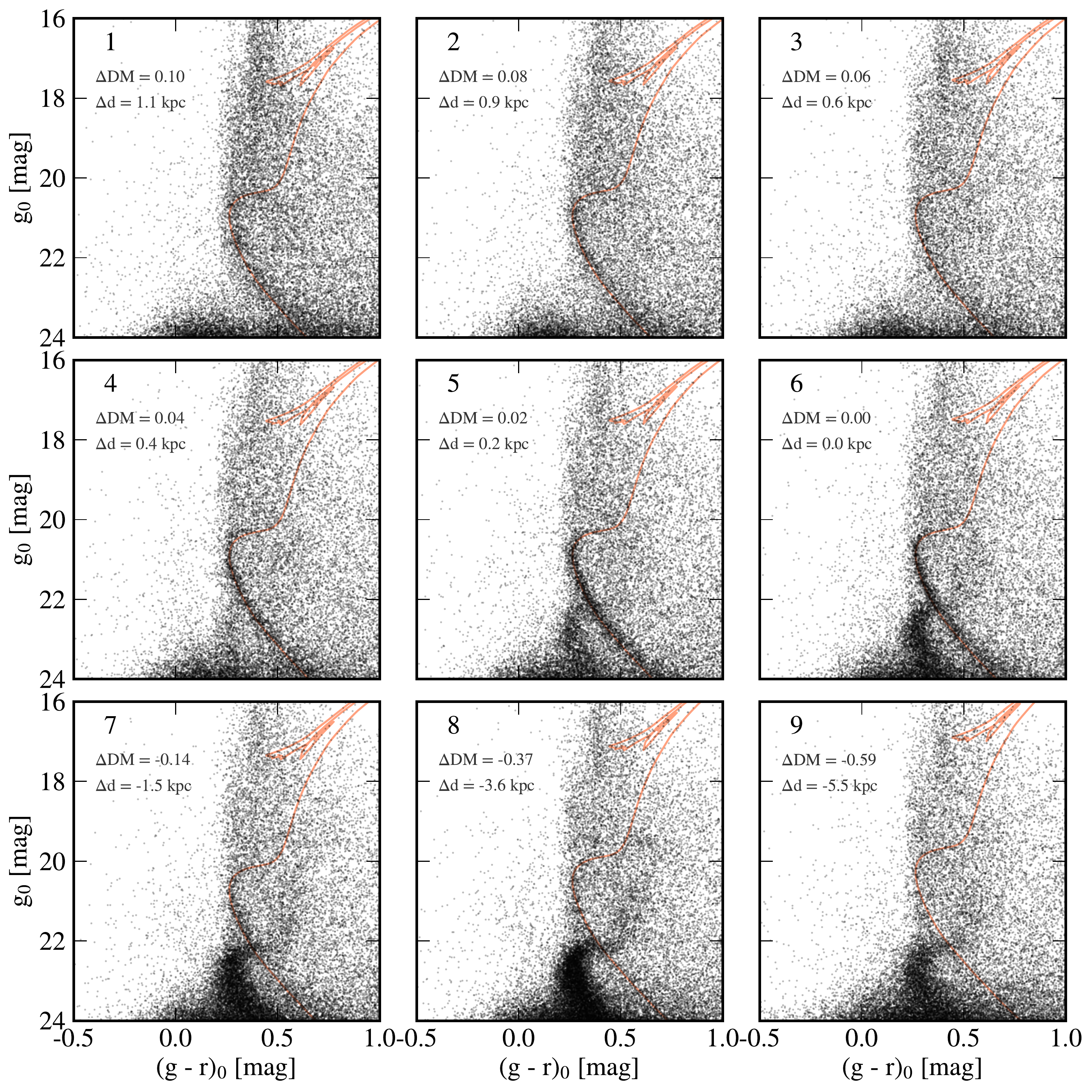}
\end{center}
\caption{
(Top) The Legacy Surveys detection of the Palomar 5 globular cluster in a coordinate system aligned with its tidal tails.
(Bottom) Color-magnitude diagrams of $\approx0.8\times3$\,deg$^2$ windows along the tails.
The regions are labeled in the top left of each panel, and their sky locations are marked in the top panel.
A stellar population consistent with Pal~5 is evident in every region, although its prominence varies between the fields.
There is a distance gradient along the stream, with the end of the trailing tail ($\phi_1\sim-15^\circ$) being the most distant, and end of the leading tail ($\phi_1\sim10^\circ$) the closest.
The fiducial Pal~5 isochrone is offset in every panel so that it matches the location of the main sequence (offsets in distance modulus and distance from the fiducial Pal~5 values are indicated in the top left).
}
\label{fig:cmds}
\end{figure*}

We study the Palomar~5 system in the photometric catalog of DECam Legacy Survey \citep[DECaLS, part of the DESI Legacy Imaging Surveys,][]{dey2019}\footnote{see also \url{www.legacysurvey.org}}.
The survey was designed to provide deep $grz$ imaging at high galactic latitudes, with the targeted $5\,\sigma$ depth of $g=24$, $r=23.4$, and $z=22.5$.
In addition to data obtained as a part of the survey, DECaLS also includes imaging from publicly available DECam data in the survey footprint.
We conducted a targeted survey of Pal~5 whose data products are now a part of DECaLS (NOAO Proposal ID nos. 2014A-0321, PI: Geha; 2014A-0611, PI: Munoz; 2015A-0620, PI: Bonaca).
The median (minimum) $5\,\sigma$ PSF depth in the Pal~5 region is $g=25.6(25.3)$, $r=25.1(24.8)$, $z=24.1(23.4)$, which makes DECaLS the deepest and largest-area survey of Pal~5.

To select likely Pal~5 stars, we first queried the DECaLS DR8 sweep catalogs\footnote{\url{http://legacysurvey.org/dr8/files/\#sweep-catalogs-region-sweep}} for point sources.
The catalog was constructed using \tractor\ forward-modeling code for source extraction\footnote{\url{https://github.com/dstndstn/tractor}} and a source was classified as \texttt{`PSF'} if the PSF model was preferred to the round exponential model used to represent galaxies\footnote{\url{https://github.com/legacysurvey/legacypipe}}.
We removed spurious sources by requiring \texttt{allmask\_g==0}, \texttt{allmask\_r==0} and \texttt{brightstarinblob==0}.
With this clean stellar catalog, we identified stars on Pal~5's main sequence in the dereddened color-magnitude diagram \citep[using the re-calibrated SFD dust map;][]{Schlegel:1998, Schlafly:2011}.
Specifically, we selected stars following an 11.5\,Gyr MIST isochrone with $\rm [Fe/H] = -1.3$ \citep{Choi:2016} between $20<g<23.7$.

In the top panel of Figure~\ref{fig:cmds} we present the sky distribution of likely Pal~5 main sequence stars.
The $(\phi_1,\phi_2)$ coordinate frame is oriented along the great circle that best-fits the Pal~5 stream, while keeping the cluster at the origin $(\phi_1 = 0^\circ,\phi_2 = 0^\circ)$  and its motion in the direction of positive $\phi_1$ \citep{gala}.
There is a distance gradient along the Pal~5 stream \citep{Ibata:2016}, so to increase its contrast against the field Milky Way stars, we applied the isochrone selection at two distances: 23\,kpc for $\phi_1<=0^\circ$ and 19\,kpc for $\phi_1>0^\circ$.
With our map, Pal~5 is continuously detected between $\phi_1=-15^\circ$ and $7^\circ$.

The color-magnitude diagrams (CMDs) of stars in different regions of the Pal~5 stream are shown in the bottom of Figure~\ref{fig:cmds}.
Each panel contains stars from a $3^\circ$ long and $0.8^\circ$ wide area marked in the top of Figure~\ref{fig:cmds} (see red boxes).
The Pal~5 main-sequence turn-off at $g\sim20.5$ stands out in all fields, but the depth to which the main sequence is detected varies from $g\sim24$ close to the cluster to $g\sim22$ in the leading tail.
The non-uniform detection depth is partly due to variable photometric depth (the coverage is shallower in the leading tail), and partly due to contamination from the Sagittarius stellar stream (main-sequence turn-off at $g\sim22$ for $\phi_1\gtrsim-5^\circ$) and faint galaxies ($g\gtrsim23$).
Despite these challenges, the Pal~5 main-sequence is evident even in field~9 ($9^\circ<\phi_1<12^\circ$), beyond the apparent leading tail truncation at $\phi_1\sim7^\circ$.
This indicates that Pal~5 tails may be longer than previously thought (\citealt{Bernard:2016}).

To improve our selection of Pal~5 members, we first employ the $z$-band to distinguish between faint stars and galaxies more efficiently.
In DECam filters, the $g-z$ color of stars bluer than $g-r\lesssim1.2$ is approximately linear with the $g-r$ color \citep[e.g.,][]{dey2019}.
We adopted the stellar locus of $(g-z)_0 = 1.7\times(g-r)_0 -0.17$.
On the other hand, galaxies span a wider distribution of redder $g-z$ colors at a fixed $g-r$ color.
To exclude faint galaxies, we restrict to sources within 0.1\,mag of the stellar locus.
We limit the sample to sources with $g<23$ to reduce inhomogeneities due to the shallow $z$-band coverage.

\begin{figure*}
\begin{center}
\includegraphics[width=\textwidth]{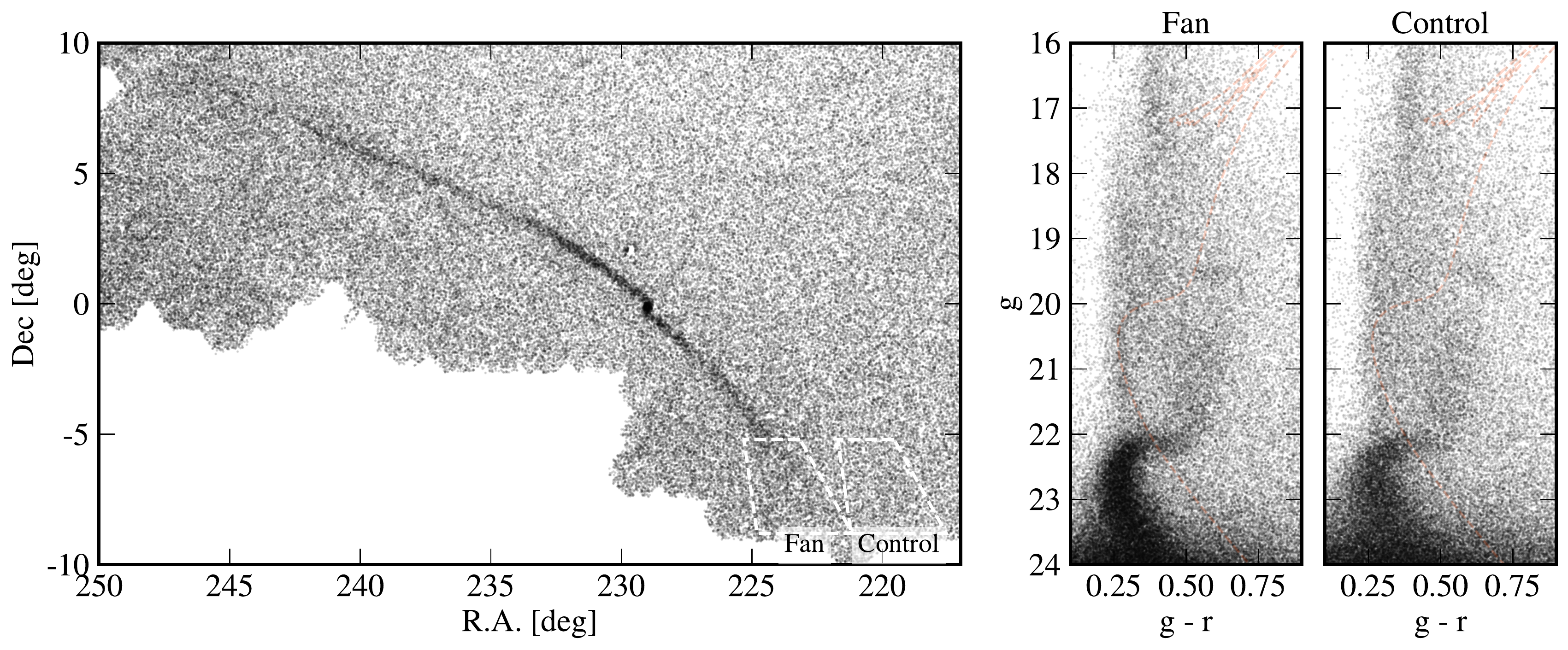}
\end{center}
\caption{
The optimized Legacy Surveys detection of the Palomar 5 system (left) reveals: (1) prominent underdensities at $\rm R.A.\approx 227^\circ$ and $235^\circ$, (2) sharp changes in the direction of the stream track at $\rm R.A.\approx 226^\circ$ and $233^\circ$, and (3) an extremely wide, low surface-brightness extension (``fan'') of the leading tail beyond $\rm Dec\lesssim-5^\circ$.
The ``fan''  has a stellar population consistent with Pal 5 (see the color-magnitude diagram comparison to a control field, right, with the Pal~5 isochrone overplotted in orange).
These features may be evidence of Pal~5's perturbed dynamical history.
}
\label{fig:map}
\end{figure*}

To further refine the Pal~5 membership, we next perform the isochrone selection that varies the distance along the stream.
We start by approximating the distance to the nine segments of the Pal~5 stream indicated in Figure~\ref{fig:cmds} from the locations of their main-sequence turn-offs.
The adopted distance modulus relative to the cluster bin (segment 6) is indicated in CMD panels at the bottom of Figure~\ref{fig:cmds}, and the corresponding isochrone is shown in orange.
Qualitatively matching the location of the turn-off, we find that the leading tail is closer than the trailing tail, in agreement with the precise RR Lyrae distance trend found in \citet{Price-Whelan:2019}.
We perform the updated isochrone selection in $2^\circ$ bins of $\phi_1$ by interpolating the location of the main-sequence selection box between these distance estimates.
With the improved star--galaxy separation and a selection that accounts for the distance gradient along the stream, Pal~5 tails are prominent between $\rm R.A.\sim223^\circ$ and $\sim245^\circ$ (Figure~\ref{fig:map}).
There is still some contamination remaining, which cannot be reduced with \gaia\ data due to the faint magnitudes of our sources.
Future surveys, e.g., \textsl{WFIRST}, will deliver precise proper motions for faint stars \citep{Sanderson:2017}, and enable additional selection of Pal~5 members based on kinematics.
At the present, however, Figure~\ref{fig:map} shows the cleanest selection of Pal~5 stars.

The DECaLS map of Pal~5 presented in Figure \ref{fig:map} reveals qualitatively new features in the stream.
First, the leading tail extends to $\rm Dec\sim-7^\circ$, beyond the previously detected edge at $\rm Dec\sim -5^\circ$ \citep{Bernard:2016}.
This newly detected extension of the leading tail is very wide ($\sigma\sim 0.25^\circ$) and in stark contrast with the trailing tail at the same distance from the cluster ($\sigma\sim 0.1^\circ$). We refer to this wide, low surface-brightness extension as the ``fan'' (\citealt{Pearson:2015}).
In the right panels of Figure \ref{fig:map}, we compare the color-magnitude diagram of the fan to an off-stream control field.
The fan has a stellar population consistent with Pal 5 (orange isochrone, see also Figure \ref{fig:cmds}), in contrast to the control field, and is therefore likely an extension of the leading tail.

Next, there are significant variations in the stellar density along the stream.
At a fixed distance from the cluster, the trailing tail is denser than the leading tail.
Furthermore, both tails feature a prominent gap in the stellar density, located at $\rm R.A.\approx 227^\circ$ and $235^\circ$ for the leading and trailing tail, respectively.
Finally, the stream track sharply changes direction at $\rm R.A.\approx 226^\circ$ and $233^\circ$.

\section{Stream density modeling}
\label{sec:densitymodel}

We quantify the variations along the Pal 5 stream and measure the stream track, width, and surface density by constructing a joint model of the stream and background stellar density.
In bins of $\phi_1$, we model the (CMD-filtered) density distribution in $\phi_2$ using a single Gaussian component for the stream and a linearly varying background \citep[similar to the density modeling described in][]{Price-Whelan:2018}.
We initialize bin centers between $\phi_1 \in (-20, 10)^\circ$ with a spacing of $h_{\phi_1} = 0.75^\circ$, but when modeling the density in $\phi_2$ in any bin, use partially overlapping bins of width $1.5\,h_{\phi_1}$.

\begin{figure}
\begin{center}
\includegraphics[width=0.8\columnwidth]{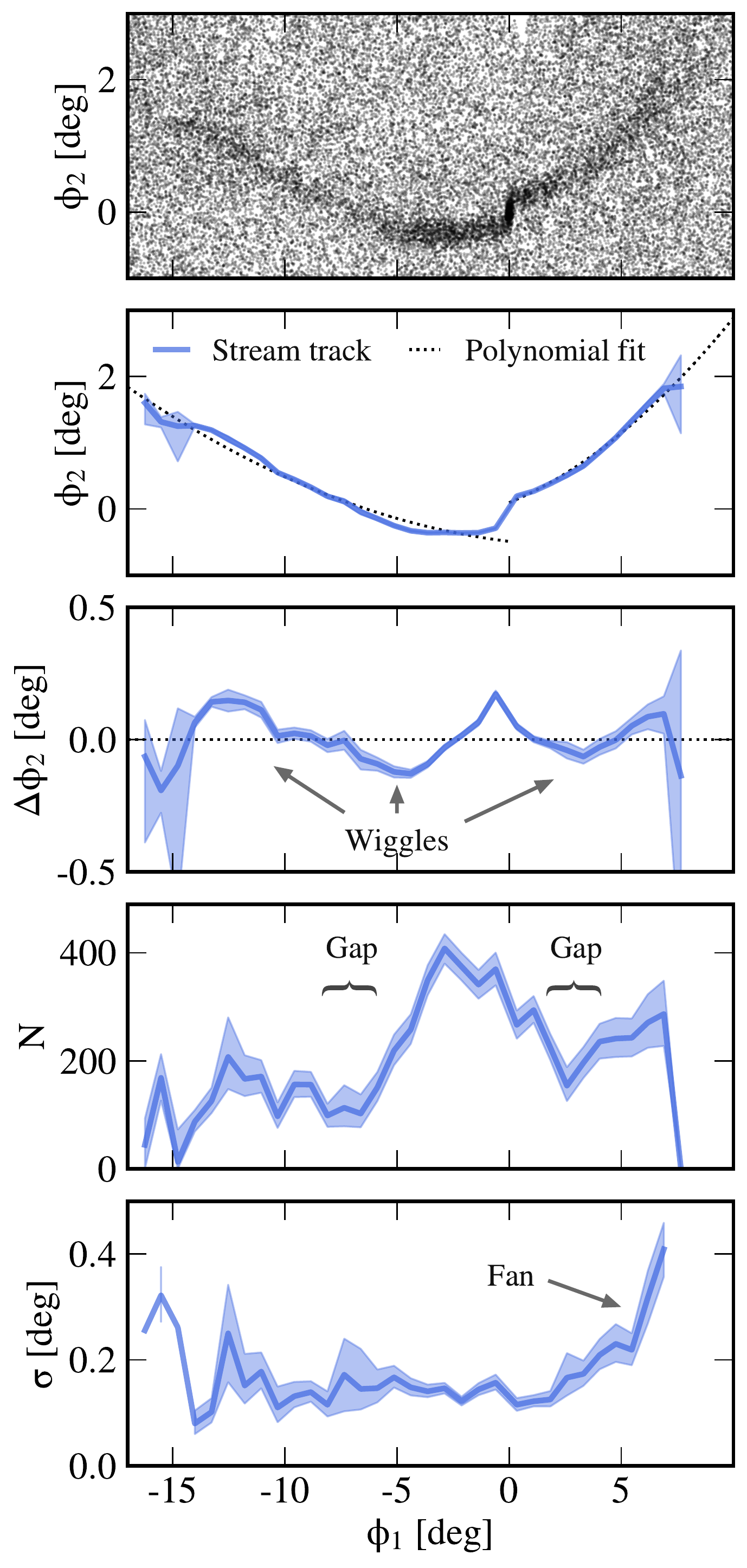}
\end{center}
\caption{
Members of the Pal~5 stream, in a coordinate system aligned with the stream, used for 1D density modeling (top row).
The inferred stream track and its residual from second-order polynomial fits to each tidal arm (rows 2 and 3, respectively) show sharp changes in the tails' direction (``wiggles'').
The density profile (row 4) is asymmetric overall, and features two prominent underdensities (``gaps'').
The trailing tail remains thin even far from the cluster, while the leading has a marked increase in width past $\phi_1\gtrsim5^\circ$ (the ``fan'', row 5).
}
\label{fig:profiles}
\end{figure}

In detail, within each $\phi_1$ bin, we construct a model for the CMD-filtered density distribution in $\phi_2$ as
\begin{align}
    p(\phi_2 \given \bs{\theta}) &=
        f \, p_{\textrm{str}}(\phi_2 \given \mu, \sigma) +
        (1-f) \, p_{\textrm{bg}}(\phi_2 \given a, b) \\
    p_{\textrm{str}}(\phi_2 \given \mu, \sigma) &=
        \mathcal{N}(\phi_2 \given \mu, \sigma^2)\\
    p_{\textrm{bg}}(\phi_2 \given a, b) &\propto
        a \, \phi_2 + b
\end{align}
where $\bs{\theta} = (f, \mu, \sigma, a, b)$ are the density model parameters, $f$ is the fraction of stars in Pal 5 in the current bin, $\mu$ is the stream centroid (track), $\sigma$ is the stream width, $a, b$ are parameters of the linear background model, and $\mathcal{N}(x \given \mu, \sigma^2)$ denotes the normal distribution over $x$ with mean $\mu$ and variance $\sigma^2$.
We assume uniform prior probability distributions, $\mathcal{U}(\alpha, \beta)$ (defined over the domain $(\alpha, \beta)$), for $(f, a, b)$, a Gaussian prior on $\mu$ to keep it close to the initialized stream track, and a prior that is uniform in $\ln \sigma$ for the stream width,
\begin{align}
    p(f) &= \mathcal{U}(f \given 0, 1)\\
    p(\mu) &= \mathcal{N}(\mu \given \mu_0, 1^\circ)\\
    p(\sigma) &= \mathcal{U}(\ln\sigma \given {-10}, 10)\\
    p(a) &= \mathcal{U}(a \given {-10}, 10)\\
    p(b) &= \mathcal{U}(b \given {-10}, 10)\\
    p(\bs{\theta}) &= p(f) \, p(\mu) \, p(\sigma) \, p(a) \, p(b)
\end{align}
where $\mu_0$ is the initialized location of the stream track in any bin.

We use an ensemble Markov Chain Monte Carlo (MCMC) sampler \texttt{emcee} \citep{Foreman-Mackey:2013} to generate samples from the posterior probability distribution over our model parameters
\begin{align}
    p(\bs{\theta} \given {\phi_2}_N) &\propto
        p(\bs{\theta}) \, \prod_n^N p(\phi_{2, n} \given \bs{\theta})
\end{align}
where ${\phi_2}_N$ are the $\phi_2$ positions of all $N$ CMD-filtered stars in the given $\phi_1$ bin.
We use 64 walkers in the sampler and, for a given $\phi_1$ bin, initialize the positions of the walkers by sampling from a small Gaussian (with dispersion $10^{-4}$) around the fiducial parameter position $\bs{\theta}_0 = (0.2, \mu_0, 0.1^\circ, 0, 1)$, with $\mu_0$ determined by binning the $\phi_2$ distribution and taking the bin center with the most stars.
We initially burn-in the samplers for 512 steps, then reset and re-run for an additional 1024 steps.
We then compute the median and (16, 84)th percentile parameter values in each $\phi_1$ bin.

Figure~\ref{fig:profiles} shows the results of running this density model on the observed distribution of Pal~5 members (top row).
In the second row, we present the inferred stream track (solid blue line and confidence interval).
As previously found by \citet{Bernard:2016}, the trailing tail is significantly longer than the leading tail ($\approx16^\circ$ vs $\approx8^\circ$ in our data).
The general curvature of the stream is well-captured by second-order polynomials (fits to the leading and the trailing arm are shown as dotted lines).
However, the stream track changes direction at several locations, as evident in the track's deviation from the polynomial fits (third row).
Two of these wiggles appear almost symmetrically in the leading and trailing tails at $\phi_1\approx3.5^\circ$ and $\phi_1\approx-4.5^\circ$, respectively, while the trailing tail has another wiggle at $\phi_1\approx-13^\circ$.

Next, we quantify the total number of stars in the Pal~5 stream, as well as its density profile (fourth row).
Accounting for the field Milky Way population, we find $3000\pm100$ Pal~5 stars between $20<g<23$ excluding the progenitor.
Integrated out to $|\phi_1|=8^\circ$ (limited by the extent of the leading arm), the trailing and the leading tail have approximately the same number of stars.
But much like the total stream extent, the density profile is also asymmetric, so within $5^\circ$ from the cluster, the trailing tail has a $50\,\%$ excess over the leading tail.
On top of these global density variations, Pal~5 features two prominent underdensities at $\phi_1\approx-7^\circ$ and $\phi_1\approx3^\circ$.
In the shallower CFHT data, these gaps were consistent with the background \citep{Erkal:2016}, while in our deeper data they are clearly detected above the background.

Finally, we measure how the width of Pal~5 varies along the stream (fifth row).
Close to the cluster, both the leading and the trailing tail are thin, $\sigma\approx0.15^\circ$.
The trailing tail remains as thin throughout, with the exception of lower-density regions, where the inference is more uncertain due to a higher level of contamination.
On the other hand, the leading tail rapidly increases in width to $\sigma\approx0.4^\circ$ (130\,pc) at $\phi_1\approx7^\circ$ (assuming a distance of 18.9 kpc, see Figure~\ref{fig:cmds}).
Since the tidal debris is fanned over a larger area, its surface density is low, and thus this part of the stream avoided detection in shallower data sets.

To summarize, we detected variations in the direction, density and width of the Pal~5 stream.
In Section~\ref{sec:sim} we explore the origin of these features by running the same density model on four different Pal~5 simulations.

\section{Simulations of Pal 5's evolution}
\label{sec:sim}
To explore plausible mechanisms that could produce the observed morphology of the Pal 5 stream, we run a suite of Pal~5 simulations.
In particular, we investigate whether Pal~5's wide leading tail can be explained due to chaotic regions in the potential, or whether torques from the Galactic bar are necessary to explain its width. We further investigate what might have caused the length asymmetry between the leading and trailing tail, as well as the gap in the trailing tail.
In Section~\ref{sec:potential}, we describe the potentials we use to simulate the evolution of Pal~5, and in Section~\ref{sec:modeling} we describe our simulation setup.
We show the results of our analyses in Section \ref{sec:sim_results}.

\subsection{Gravitational potentials}
\label{sec:potential}
We simulate the evolution of Pal~5 in two classes of three-component Galactic potentials implemented in the  \package{gala} package \citep{gala}:

\begin{itemize}
\item[1.] {\bf Static potentials}: we use the \texttt{MWPotential2014} \citep{Bovy:2015} consisting of a Miyamoto-Nagai disk (\citealt{Miyamoto:1975}), a bulge modeled as an exponentially cut off, power-law density profile, and an axisymmetric NFW dark matter halo (\citealt{Navarro:1996}).
We set the flattening of the NFW halo either to the canonical value, $q_z = 0.94$, or to an unphysical $q_z = 0.5$ to investigate Pal~5's morphology on a regular as well as chaotic orbit, respectively (see Section \ref{sec:modeling}).

\item[2.] {\bf  Barred potentials}: we use the same disk and halo as in the \texttt{MWPotential2014} \citep{Bovy:2015}, but include a Galactic bar instead of a bulge.
Following \citet{wang:2012}, we compute the barred potential as a basis-function expansion (BFE) representation of a triaxial, exponential density profile:
\begin{align}
    \rho_{\rm bar} &= \rho_0 [{\rm exp} (-r^2_1/2) + r_2^{-1.85} {\rm exp}(-r_2) ]\\
    r_1 &= \left[\left((x/x_0)^2 + (y/y_0)^2\right)^2 +( z/z_0)^4\right]^{1/4}\\
    r_2 &= \left[\frac{q^2(x^2 + y^2) + z^2)}{z_0^2}\right]^{1/2}
\end{align}
where the scale lengths are $x_0$ = 1.49 kpc, $y_0$ = 0.58 kpc, $z_0$ = 0.4 kpc, and q = 0.6. We include terms up to $n=6$, $l=6$ in the ``self-consistent field'' \citep{Hernquist:1992} BFE formalism.\footnote{\citet{Banik:2019} found that using higher order terms (e.g., n=9, l = 19) for the basis function expansion yields a better representation of the density of the bar. However, this difference does not appreciably change the morphology or kinematics of our simulated streams.}
We explore barred potentials with pattern speeds of $\Omega_b = 30$--$60~\kmskpc$ in increments of $0.5~\kmskpc$, and we test bar masses of $\textrm{M}_{\rm bar} = 5 \times 10^{9}~\msun$ and $\textrm{M}_{\rm bar} = 1 \times 10^{10}~\msun$ \citep{Portail:2017}.
We fix the present day angular offset from the Galactic x-axis in the direction of rotation to $\alpha = 27^\circ$.
\end{itemize}

Note that \citet{wang:2012} constructed a bar with a pattern speed of $\Omega_b$ =  60 $\kms$ kpc$^{-1}$, which has a co-rotation radius, $r_{\rm CR} = 3.7$ kpc.
In this paper, however, we explore a range of pattern speeds, which will lead to different co-rotation radii for different pattern speeds.

As bars are not expected to extend much beyond their co-rotation radius \citep[e.g.,][]{weiner:1999, Debattista:2002, Debattista:2002b}, we adjust the physical scaling of the bar when we vary the pattern speed.
In particular, we compute the co-rotation radius, $r_{\rm CR}$, for the mass profiles of the static potential for any given pattern speed, $\Omega_b$.
We then scale the bar potential for a given pattern speed, $\Omega_b$:
\begin{equation}\label{eq:scale}
r_{s, \Omega_b}  = r_{{\rm CR}, \Omega_b}/r_{{\rm CR, Wang 2012}}
\end{equation}
If the scaling in Equation \ref{eq:scale} is not included \citep[as in, e.g.,][]{Pearson:2017, Erkal:2017, Banik:2019}, the gravitational influence of the bar at a fixed radius will be too strong (weak) for faster (slower) pattern speeds, respectively.

\subsection{Stream simulation setup}
\label{sec:modeling}

To simulate the evolution of Pal~5 in a given potential model, we first fit the stream track of Pal 5 in the static potential---that is, we fit for Pal 5 initial conditions that generate mock stream models that fit the observed sky track of the stream (Section~\ref{sec:densitymodel}).
We use the observed 6D phase space coordinates of the Pal 5 cluster (sky positions from \citealt{Odenkirchen:2002}, distance from \citealt{Harris:2010}, radial velocity from \citealt{Kuzma:2015}, and proper motion from \citealt{Fritz:2015}, see also Table~1 in \citealt{Pearson:2017}) to initialize the track fit and transform our fitted 6D phase space coordinates into a Galactocentric frame by assuming $v_{lsr} = (11.1, 24.0, 7.25) ~\kms$,  $v_{circ} = 220~\kms$, and a distance from the Sun to the Galactic center of 8.1 kpc (\citealt{Schonrich:2010}, \citealt{Schonrich:2012}).
The best-fit present-day kinematics of Pal 5 that reproduce the observed sky track in the static potential are $\alpha = 229.0264^\circ$, $\delta = -0.1368^\circ$, $d = 22.5~\kpc$, $v_r = -56.2~\kms$, $\mu_{\alpha\,\cos\delta} = -2.21~\masyr$, and $\mu_\delta = -2.23~\masyr$.

With the present-day location of Pal~5 in hand, we generate the orbit of the progenitor system by integrating Pal~5's current 6D position back in time for 3\,Gyr in steps of 0.5\,Myr.
Starting from this point and integrating forward to the present day, we simulate the cluster's disruption and the formation of the stream by releasing two particles from the progenitor's Lagrange points at every time step, using the \citet{Fardal:2015} distribution function implemented in the \package{gala} package \citep{gala}.
We account for the self-gravity of the Pal~5 cluster by including a Plummer mass component with the initial half-mass radius of $r_h = 4~\textrm{pc}$ and the initial effective mass, $\textrm{M}_p = 14,404~\msun$.
This mass was obtained from the stream track fitting and sets the scaling between the Lagrange points for the modified ``particle spray'' method (\citealt{Fardal:2015}).
This is comparable to $12,200 \pm 200~\msun$ that \citet{Ibata:2017} report for the present-day mass of the cluster and tails combined.

To investigate Pal~5's evolution on a regular orbit, we first run a ``particle-spray'' simulation in the static potential, setting $q_z = 0.94$ \citep{Bovy:2016}.
Subsequently, we run the same simulation setting $q_z = 0.5$ to place Pal 5 on a strongly chaotic orbit.
We then simulate the evolution of the stream in time-dependent barred potentials, varying the pattern speed of the bar while updating its physical scaling (see Section~\ref{sec:potential}) and varying the mass of the bar as described above.

\subsection{Simulation results}
\label{sec:sim_results}
\begin{figure*}
\begin{center}
\includegraphics[width=\textwidth]{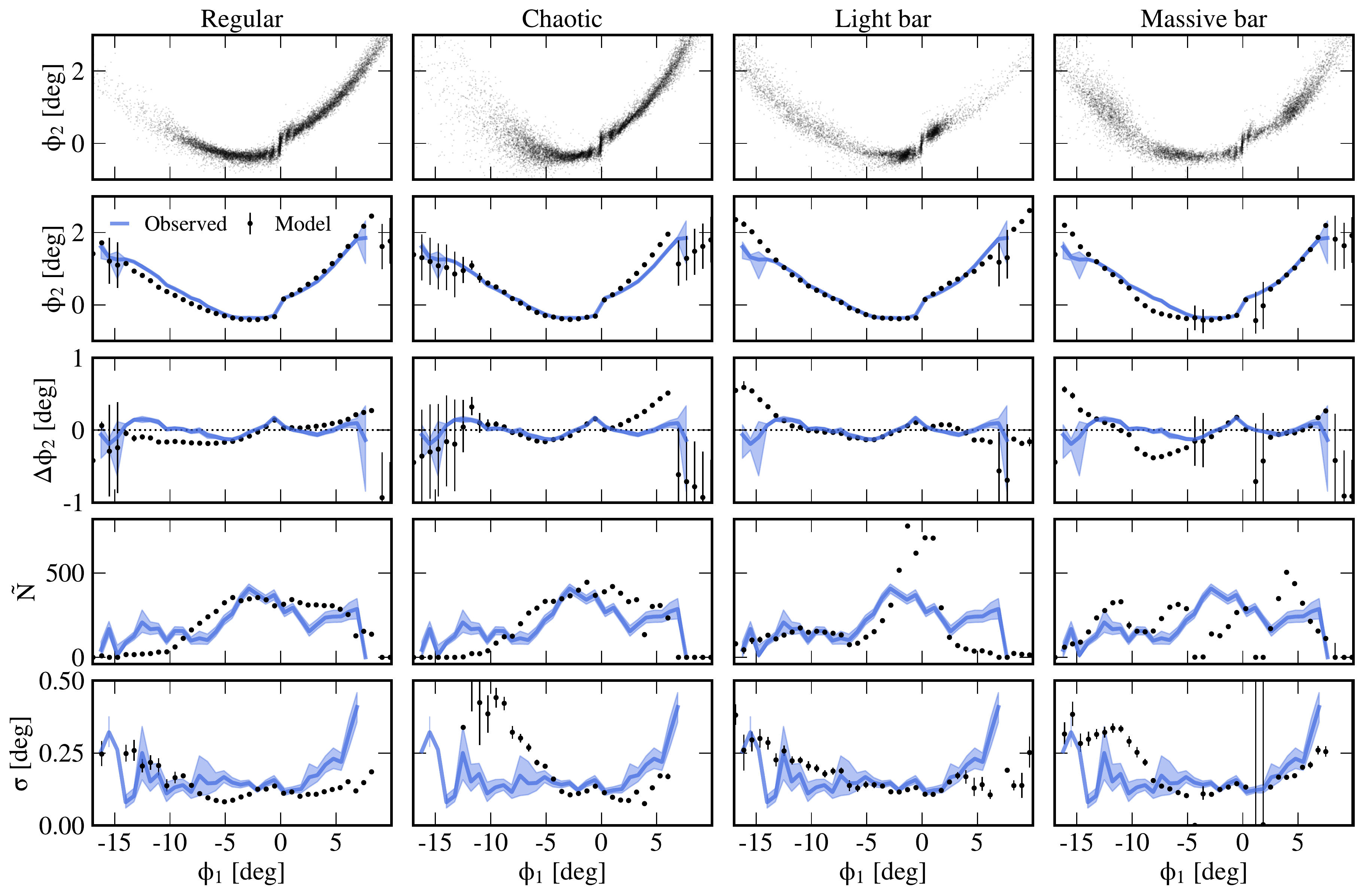}
\end{center}
\caption{
Properties of four Pal~5 simulations: ``Regular'' (static potential and a regular orbit), ``Chaotic'' (static potential and a chaotic orbit), ``Light bar'' (time-dependent potential with a M$_{\rm bar} = 5\times 10^{9}\,\msun$, $\Omega_b = 38\,\kms$ kpc$^{-1}$ bar), and ``Massive bar'' (M$_{\rm bar} = 1\times 10^{10}\,\msun$, $\Omega_b = 45\,\kms$ kpc$^{-1}$).
The top row shows the spatial distribution of tracer particles, while rows 2 through 5 present the derived 1D properties along the stream (black points): $\phi_2$ positions of the mean stream track, track deviations from a second order polynomial, number density and width, respectively.
These simulations individually reproduce many of the features observed in Pal~5 (blue lines, reproduced from Figure~\ref{fig:profiles}), but no model matches all of the features simultaneously.
}
\label{fig:sims}
\end{figure*}

In Figure~\ref{fig:sims} we summarize the results of our suite of Pal~5 simulations.
In the top row we present the morphology of the simulated Pal~5 mock-streams evolved in the static potentials with two different flattenings ($q_z = 0.94$ and $q_z = 0.5$) and we show two examples of mock-streams evolved in the ``light'' (M$_{\rm bar} = 5\times 10^9~\msun$) and ``massive'' (M$_{\rm bar} = 1\times 10^{10}~\msun$) time-dependent barred potentials.
All mock-streams are visualized in the coordinate system aligned with Pal 5's tidal tails (see also Figure~\ref{fig:cmds}), and the leading tail is at positive $\phi_1$.
The four bottom rows of Figure~\ref{fig:sims} show results of the 1D density model for the stream track (second row), the track deviations from a second order polynomial (third row), stream number density (fourth row), and the stream width (fifth row) of the simulated mock-streams (black) as well as of the data (blue shaded region, same as in Figure~\ref{fig:profiles}).

We used the mildly flattened ($q_z = 0.94$) static potential to fit the track of the stream (see Section \ref{sec:modeling}), and it is therefore not surprising that the track (second row) and track deviations (third row) of the mock-stream evolved on the ``Regular'' orbit (left column, black) looks similar to the track of the data (blue).
While the data shows variability in the number density of stars (fourth row), the mock-stream, however, has a similar number of stars in both the trailing and the leading tail.
Additionally, the width of both the leading and the trailing mock-stream are too narrow for |$\phi_1| < 7$ as compared to the data (fifth row).
While the overall width trends for the trailing tail  at larger negative $\phi_1$ are reproduced, the leading tail of the simulated stream is too narrow as compared to the widened ``fan'' seen in the leading tail of the data.

For the mock-stream evolved in a flattened potential, we chose a flattening of $q_z = 0.5$  in order to investigate the morphology of a mock-stream on a chaotic orbit (see also \citealt{Fardal:2015}).
The track of the leading tail (third row, black) deviates from the observed stream track (blue) reflecting the extreme flattening of the halo shape.
Interestingly, the number density of stars is roughly symmetric for the leading and trailing tail  (fourth row), which was also the case for the regular orbit.
While the width (fifth row) of the leading tail is slightly wider for the chaotic mock-stream than for the mock-stream evolved on a regular orbit (left), the mock-stream width is still too narrow when compared to the leading tail data (blue).
The trailing tail, on the other hand, is much wider for the chaotic mock-stream than for the data.
Thus, the chaotic orbit induces a widening of both the leading and trailing tail, but this widening does not mimic the particular asymmetric widening of the data (see also \citealt{Pearson:2015}, \citealt{Price-Whelan:2016}).

We selected two specific examples of mock-streams evolved in barred potentials (see ``light bar'' and ``massive bar'' columns in Figure \ref{fig:sims}) through visual inspection of the morphology, track and width of the simulated streams for streams evolved in the time-dependent potentials with bar pattern speeds ranging from $\Omega_b = 30$--$60~\kmskpc$.
The ``light bar''  ($\textrm{M}_{\rm bar} = 5\times 10^{9}~\msun$, $\Omega_b = 38 ~\kmskpc$) mock-stream simulation demonstrates a scenario in which the leading tail has been abruptly truncated (see also \citealt{Price-Whelan:2016b}, \citealt{Pearson:2017}, \citealt{Erkal:2017}).
This truncation is apparent in both the morphology (top row), number density of stars (fourth row), as well as in the width of the mock-stream (fifth row).
In particular, the number of stars in the leading tail at $\phi_1 > 2.5$ is low when compared to the data (blue).
The width of the trailing mock-stream matches the observed data well, and the part of the leading tail that has not been truncated also follows the overall trends of the data.

The mock stream evolved in a potential including the  ``massive bar'' (M$_{\rm bar} = 1\times 10^{10}~\msun$, $\Omega_b = 45~\kmskpc$), demonstrates a scenario in which the leading tail has been widened by the bar (fifth row), to a similar width as observed in the data.
In this scenario, the leading tail has not been truncated and there is a comparable number of stars in the leading and trailing tail.
The bar has, however, induced over- and under-densities (``gaps'') throughout both the leading and trailing tail (fourth row).
While this is also seen in the data, the mock-stream visualized here does not follow the gaps seen in the data exactly (located at $\rm R.A.\approx 227^\circ$ and $235^\circ$ for the leading and trailing tail, respectively, which is equivalent to $\phi_1 \approx 2.5^\circ$ and $-7^\circ$).
Interestingly, for this particular pattern speed, the bar has shifted the stream track of the trailing tail  (see second and third row).
This effect is apparent in the data on small scales where the track also appears to be ``wiggling''.
Thus, this mock stream demonstrates a scenario in which both the leading and trailing tails have been perturbed by the bar, and where the leading tail width is similar to the width of the data.

While none of the above mock-stream simulations reproduce all aspects of the data simultaneously, they do provide insight into the mechanisms that could be perturbing the stream.
In the next Section we discuss the implications of our results.

\section{Discussion}
\label{sec:discussion}
We presented deep $grz$ photometry of the Palomar~5 tidal tails from the Legacy Surveys catalog that enables the cleanest selection of Pal~5 members to date. This detailed view has revealed Pal~5 as a complex system, adding to the numerous examples of stellar streams that show signs of perturbations (e.g. \citealt{Sesar:2016}, \citealt{Price-Whelan:2018}, \citealt{Bonaca:2019a}).
One-dimensional modeling of the Pal 5 debris' spatial distribution revealed significant changes in the width and density along the tidal tails, as well as changes in their direction.
Stream simulations in different gravitational potentials separately capture many of the features observed in Pal~5, including gaps and an asymmetry in the length and the width of the leading and trailing tail.
However, these simulations fail to reproduce the specific observed gaps in the stream density and the specific wiggles of its track.
In this section we discuss perturbing mechanisms that could jointly explain this transformed view of the Pal~5 stream.

The striking asymmetry in the length and width of Pal~5's two tidal arms is immediately obvious from Figure~\ref{fig:map}.
Recently, \citet{Starkman:2019} reported an extension of Pal~5's leading tail that nearly matches the extent of the trailing tail based on a selection of main sequence turn-off stars that have proper motions similar to the Pal~5 cluster.
This choice of sparse tracers limits the confidence in estimating both the width and the density along the tails.
So while our deep photometry also indicates the leading tail extending further (Figure~\ref{fig:map}), this extension (i.e., the fan) has a significantly larger width and lower surface density.
At a fixed density threshold, we confirm the original \citet{Bernard:2016} discovery that Pal~5's leading tail is only half as long as the trailing tail.

\citet{Pearson:2017} showed that Pal~5's short leading tail may be truncated through a perturbation by the Milky Way's Galactic bar.
This scenario predicts that the bar sweeps tidal debris from the leading tail to a much wider area.
Deep, wide-field DECaLS photometry enabled us to trace the leading tail as it fans out to $\rm Dec\sim-7^\circ$, while the color-magnitude diagram implies Pal~5 debris has been spread out to even lower surface-brightness level between $-7^\circ\gtrsim\rm Dec\gtrsim-10^\circ$.
In Section \ref{sec:sim_results}, we showed that this low surface-brightness feature can indeed be induced if Pal~5 has been perturbed by the bar.

The bar is a prominent perturber that affects objects within 5\,kpc from the Galactic center \citep[e.g., the Ophiuchus stream,][]{Price-Whelan:2016b, Hattori:2016} to the Solar circle and beyond \citep[e.g., local phase-space overdensities][]{Hunt:2018, Monari:2019}.
Recent progress investigating the Galactic bar, is converging on a pattern speed close to $\Omega_b \sim 40 ~\kms$ kpc$^{-1}$ \citep[e.g.,][]{Clarke:2019, Sanders:2019, Bovy:2019}, however the exact bar properties remain somewhat uncertain.
A combination of future deeper imaging along Pal~5's leading tail and a quantitative analysis of its width and density could provide important constraints on the mass and pattern speed of the Galactic bar.

In addition, streams can also be perturbed by smaller-scale objects, such as dark-matter subhalos, spiral arms and molecular clouds \citep[e.g.,][]{Yoon:2011,Amorisco:2016, Banik:2019}.
An encounter with these more compact perturbers produces a gap in a stellar stream \citep[e.g.,][]{Johnston:2002,Ibata:2002}, and tidal tails of Pal~5 have long been searched for such density variations.
The findings so far have been conflicting, with the number of gaps reported in Pal~5 ranging from five \citep{Carlberg:2012} to no gaps \citep{Ibata:2016}.
With DECaLS photometry, we confirm the large-scale density variations reported by \citet{Erkal:2017}: the two most prominent underdensities in Pal~5 are a $\sim5^\circ$ gap in the trailing and a $\sim1^\circ$ gap in the leading tail.
Based on mass arguments, \citet{Erkal:2017} suggested that the large gap in Pal~5 originates from a dark-matter subhalo encounter, while the small gap may have been produced by a molecular cloud.
In addition to the perturber's mass, details of the gap profile also depend on the time since the encounter and its impact parameter \citep{Erkal:2015}.
Our dataset enables a confident measurement of the gap profile above the Milky Way field contamination, which will help in disentangling different encounter parameters, and ultimately determining the origin of these perturbations.

The depth of the DECaLS catalog also allows for a more confidently determined stream track and reveals its surprising deviations: a global change in Pal~5's curvature $\sim5^\circ$ from the cluster, and $\sim20'$ oscillations, or wiggles, in the trailing tail (Figure~\ref{fig:profiles}).
While different simulations capture some of these features (the curvature in the leading tail is reproduced in the ``Chaotic'' and ``Light bar'' models, and the ``Regular'' and ``Massive bar'' models match the curvature in the trailing tail), no model recovers the whole stream track.
Fisher information considerations in a static gravitational potential imply that the stream track encodes the enclosed mass at the current location of the stream \citep{Bonaca:2018}.
The change in the curvature of Pal 5's tidal arms occurs at a nearly symmetric position along the leading and the trailing arm, so this may indicate a sudden change in the enclosed mass inside Pal~5's orbit.
For example, when a stream encounters a massive object, its overall orbit can change and produce a misalignment between the proper motions and the stream track \citep{Erkal:2018, Koposov:2019}.
Measuring the relative angle between the proper-motion vector and the stream track along the Pal~5 tails would therefore test whether the change in their curvature originates from an encounter with a massive perturber.

Similarly, the small-scale wiggle in the trailing tail may be another signature of an impact, especially as it coincides with a very prominent gap (Figure~\ref{fig:profiles}).
During the encounter, nearby stream stars receive a velocity kick that changes their orbital energies and opens a gap in the stellar stream \citep[e.g.,][]{Erkal:2015b}.
In addition, stream stars affected by the perturber can be displaced from the original stream track \citep{Bonaca:2018b}.
Simultaneous fitting of the Pal~5 stream track and its density profile could determine the origin of the trailing tail gap.
Should the encounter scenario be confirmed, track wiggles may put additional constraints on the impact geometry.

More generally, our finding that deep photometry alone can be leveraged to cleanly map tidal debris has important implications for future studies of stellar streams.
Within the Milky Way, this enables detailed mapping of streams that reside outside of \gaia's scope \citep[cf.][]{Ibata:2019}, and therefore exploring the level of perturbation in the outer halo.
Because streams at greater Galactocentric radii are less likely to have been affected by baryonic structures \citep[e.g.,][]{Banik:2019}, they are especially valuable as more robust tracers of dark-matter subhalos.
Outside of the Milky Way, stellar streams have been detected almost exclusively photometrically \citep[e.g.,][]{Martinez-Delgado:2010, Kado-Fong:2018}, and future photometric surveys are primed to discover numerous low-mass streams \citep{Pearson:2019}.
Our results suggest that dynamical inferences about the bar, spiral arms, molecular clouds, and dark-matter subhalos from stream perturbations may no longer be limited to the Milky Way, but instead expanded to large samples of galaxies.

\software{
    \package{Astropy} \citep{astropy, astropy:2018},
    \package{dustmaps} \citep{dustmaps},
    \package{gala} \citep{gala},
    \package{IPython} \citep{ipython},
    \package{matplotlib} \citep{mpl},
    \package{numpy} \citep{numpy},
    \package{scipy} \citep{scipy}
}

\acknowledgements{
It is a pleasure to thank Charlie Conroy's and Doug Finkbeiner's groups at the CfA for helpful suggestions.
This project was developed in part at the 2019 Santa Barbara Gaia Sprint, hosted by the Kavli Institute for Theoretical Physics at the University of California, Santa Barbara.
This research was supported in part at KITP by the Heising-Simons Foundation and the National Science Foundation under Grant No. NSF PHY-1748958.
The Flatiron Institute is supported by the Simons Foundation.

The Legacy Surveys consist of three individual and complementary projects: the Dark Energy Camera Legacy Survey (DECaLS; NOAO Proposal ID \# 2014B-0404; PIs: David Schlegel and Arjun Dey), the Beijing-Arizona Sky Survey (BASS; NOAO Proposal ID \# 2015A-0801; PIs: Zhou Xu and Xiaohui Fan), and the Mayall z-band Legacy Survey (MzLS; NOAO Proposal ID \# 2016A-0453; PI: Arjun Dey). DECaLS, BASS and MzLS together include data obtained, respectively, at the Blanco telescope, Cerro Tololo Inter-American Observatory, National Optical Astronomy Observatory (NOAO); the Bok telescope, Steward Observatory, University of Arizona; and the Mayall telescope, Kitt Peak National Observatory, NOAO. The Legacy Surveys project is honored to be permitted to conduct astronomical research on Iolkam Du'ag (Kitt Peak), a mountain with particular significance to the Tohono O'odham Nation.

The Legacy Surveys imaging of the DESI footprint is supported by the Director, Office of Science, Office of High Energy Physics of the U.S. Department of Energy under Contract No. DE-AC02-05CH1123, by the National Energy Research Scientific Computing Center, a DOE Office of Science User Facility under the same contract; and by the U.S. National Science Foundation, Division of Astronomical Sciences under Contract No. AST-0950945 to NOAO.

\bibliographystyle{aasjournal}
\bibliography{pal5fan}

\end{document}